\documentclass[conference
]{IEEEtran}

\usepackage[square,numbers,sort&compress]{natbib}
\usepackage{hyperref} % 
\usepackage% 
{hyperref} % 
\hypersetup{
    colorlinks,
    linkcolor={blue!80!black},% 
    citecolor={green!50!black},
    urlcolor={blue!80!black}
}

\usepackage{amsthm}
\usepackage{amsmath}
\usepackage{mathrsfs}
\usepackage{amssymb} 
\usepackage{stackrel}
\usepackage{mathtools}

\usepackage{physics}

\usepackage{verbatim}
\usepackage{enumerate}

\usepackage{slashbox} 

\usepackage[T1]{fontenc}
\usepackage{bbm} 
\usepackage{bbding}
\usepackage{keystroke}
\usepackage{pifont}
\usepackage{relsize} % 

\usepackage{graphicx}
\usepackage{xcolor}
\usepackage{tikz}
\usetikzlibrary{calc}
\usepackage{tabularx}
\usetikzlibrary{arrows,shapes}

\renewcommand{\trace}{\mathrm{Tr}}

\newcommand{\identity}{\mathbbm{1}}

\theoremstyle{remark}	\newtheorem{theorem}{Theorem}
\theoremstyle{remark}	
\theoremstyle{remark}	
\theoremstyle{remark}	
\theoremstyle{remark} \newtheorem{definition}{Definition}
\theoremstyle{remark} \newtheorem{remark}{Remark}
\theoremstyle{remark} 

\usepackage{balance}
							
\begin{document}

\title{Semantic Security with Unreliable\\ Entanglement Assistance: Interception and Loss}

\author{
    \IEEEauthorblockN{Meir Lederman and
    Uzi Pereg} \\
		\vspace{0.2cm}
    \IEEEauthorblockA{\normalsize
		\IEEEauthorrefmark{1}Faculty of Electrical and Computer Engineering, Technion \\
		\IEEEauthorrefmark{2}Helen Diller Quantum Center, Technion \\
    Email: {\tt 
 meirlederman@campus.technion.ac.il,     uzipereg@technion.ac.il}
}} 

\maketitle

\begin{abstract}
Semantic security is considered with unreliable entanglement assistance, due to one of two reasons: Interception or loss. % 
We consider two corresponding models. 
In the first model, Eve may intercept the entanglement resource. In the second model, Eve is passive, and the % 
resource may dissipate to the environment beyond her reach. We derive achievable rates for both models, subject to a  maximal error criterion and semantic security. As an example, we consider the amplitude damping channel. 
Under interception, % 
time division is not necessarily possible, and the boundary of our achievable region is disconnected.
In the passive model, % 
our rate region outperforms  time division. % 
\end{abstract}

\section{Introduction}
Information theoretic security has traditionally relied on % 
weak and strong secrecy metrics. % 
 However, from a cryptographic standpoint, these % 
 were considered inadequate \cite{bellare3160polynomial% 
 }, due to % 
 the assumption of messages being randomly and uniformly distributed. Nonetheless, real-world messages often originate  from structured data with low entropy, such as files or votes \cite{GoldfeldCuffPermuter:16p}. Semantic security has thus become the gold standard in the cryptography community \cite{bellare2012semantic, boche2022mosaics, hayashi2007upper}. % 
Semantic security ensures that the eavesdropper gains no information, without making any assumptions on  the message distribution. 
It can also be formulated as message indistinguishability, where Eve is unable to distinguish between any pair of messages \cite{boche2022semantic}.

Entanglement resources are useful in many % 
applications, including   % 
physical-layer security 
\cite{YLLYCZRCLL:20p,% 
ZlotnickBashPereg:23c}, and  % 
can significantly increase   throughput % 
\cite{% 
QiSharmaWilde:18p, notzel2020entanglement}. % 
Unfortunately, it % 
is a fragile resource % 
\cite{% 
campbell2008measurement}.
In order to generate entanglement assistance in optical communication, the transmitter first prepares an entangled pair % 
locally, and then transmits half of it % 
\cite{YCLRLZCLLD:17p}. Since % 
photons are easily  lost to the environment % 
\cite{czerwinski2022statistical},
current implementations incorporate % 
a back channel  to notify the transmitter in case of a % 
failure, with numerous repetitions. % 
This approach has clear disadvantages % 
and may even result in % 
system 
collapse. 
However, ensuring resilience and reliability is % 
critical for  developing future communication 
networks \cite{FB22}.

Communication with unreliable entanglement assistance was recently introduced in \cite{pereg2023communication} as a setup where a back channel and repetition are not required.
Instead, the % 
rate is adapted to the availability of entanglement assistance. % 
 The principle of operation % 
 ensures reliability by design. % 
Uncertain cooperation was originally studied in classical multi-user information theory  \cite{% 
HuleihelSteinberg:17p}, motivated by the engineering aspects of modern % 
networks. 
The quantum model  involves a point-to-point quantum channel and  unreliable correlations \cite{pereg2023communication,pereg2023communication-1b}.

The secrecy capacity of a quantum wiretap channel
has been investigated in various settings \cite{dupuis2013locking, fawzi2013low, boche2017classical}. 
Cai  \cite{cai2004quantum} and Devetak \cite{devetak2005private} established 
a regularized capacity  formula % 
without entanglement assistance. Boche et al. \cite{boche2022semantic} presented explicit constructions of semantic secuirty codes. % 
Qi et al. \cite{QiSharmaWilde:18p} considered secure communication with entanglement assistance.
The authors have recently considered
 communication with unreliable entanglement assistance and strong secrecy \cite{Lederman2024Secure}, 
 assuming that the information is uniform.

Here, we consider two semantic security settings of a quantum wiretap channel with unreliable entanglement assistance.
Before communication begins, the legitimate parties try to generate entanglement assistance. To this end, Alice prepares an entangled pair % 
locally and transmits one particle. % 
The particle may fail to reach Bob due to one of two reasons:
\begin{enumerate}[1)]
\item
 \emph{Interception:} While the particle travels from the transmitter, Eve tries to steal it. % 
\item
\emph{Loss:}
The particle is lost to the environment.
Yet, Eve is passive and does not gain access to the % 
resource.
\end{enumerate}
In the optimistic case, Alice and Bob generate entanglement  successfully   prior to the transmission of information. Hence, Bob can decode the information while using the entangled resource, which is not available to Eve.
However, in the pessimistic case, % 
Bob must decode without it.  
Nonetheless, % 
secrecy needs to be maintained, whether Bob, Eve, or a neutral environment hold the entangled resource.

Alice % 
encodes two messages at rates $R$ and $R'$, unaware of % 
whether Bob  holds the entanglement resource or not. 
Whereas, Bob and Eve know whether the resource is in their possession.
 In practice, this is realized through heralded entanglement generation \cite[Remark 2]{pereg2023communication}.
If the entangled resource is not available to Bob, then he decodes the first message alone; hence, the transmission rate is $R$.
Whereas, given entanglement assistance, Bob decodes both messages, hence the overall rate is $R+R'$.
The rate $R$ is thus associated with information that is guaranteed to be sent, while $R'$ with % 
the excess information that entanglement assistance provides.
In this manner, we adapt the transmission rate to the availability of entanglement assistance, while communication does not break down when the assisting resource is absent.

We establish an % 
achievable rate region for communication with unreliable entanglement assistance and semantic security, for each setting.
To demonstrate our results, we consider the amplitude damping channel. % 
In the interception model,
we encounter a phenomenon that is somewhat rare in network information theory \cite{BocheNotzel:14p1}: Time sharing is impossible and 
the boundary of our achievable 
region is disconnected.
Whereas, in the passive model, our achievable rate region outperforms time division.

In the analysis, we introduce a novel proof technique 
for the maximal error and security analysis. 
Our technique modifies the methods by Cai \cite{cai2014maximum} 
 for multiple access channels with correlated transmitters (see also \cite{ pereg2023multiple}).

\section{Coding Definitions}% 
\label{section_definitions_interception}

\begin{figure}[tbp]
\centering
        \includegraphics[scale=0.5, trim={3.5cm 9cm 0 10.25cm}, clip]{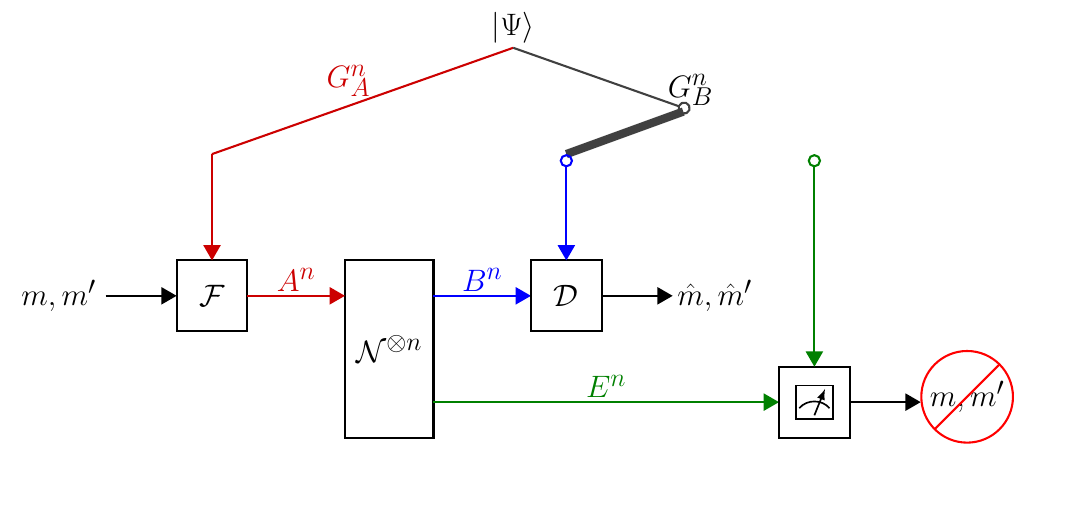} % 
\\ \vspace{-0.5cm}
(a)
\\
        \includegraphics[scale=0.5, trim={3.5cm 9cm 0 9.75cm}, clip]{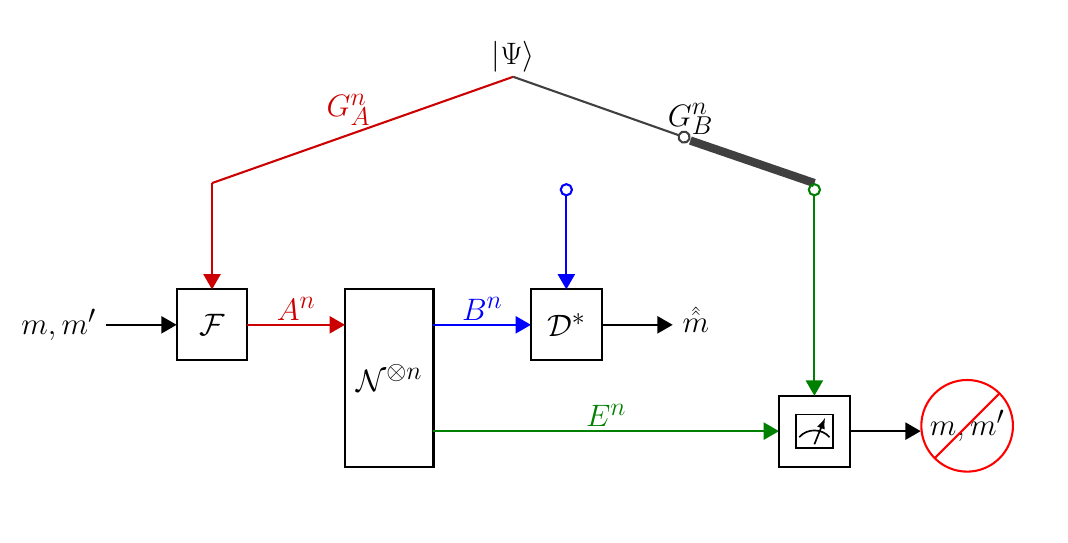} % 
\\ \vspace{-0.5cm}
(b)
    
    \caption{Interception. % 
    As Eve may steal the  resource, % 
    there are two scenarios: (a) "Left": Bob % 
    decodes both $m$ and $m'$. % 
    (b) "Right": Bob decodes $m$ alone.  % 
    }
    \label{figure_switch_Interception}
\end{figure}

Before communication begins, the legitimate parties try to generate entanglement assistance. 
In the optimistic case, % 
Alice and Bob have % 
entanglement resources, $G_A^{n}$ and $G_B^{n}$, respectively
(see Figure~\ref{figure_switch_Interception}(a)).
However, $G_B^{n}$ is not necessarily available to Bob, due to either interception or loss. % 

In the communication phase, Alice sends $n$ inputs through a memoryless quantum wiretap channel $\mathcal{N}_{A\to BE}% 
$, while she is   % 
unaware of whether Bob has % 
the entanglement resource. 
Nevertheless, based on the common use of heralded entanglement generation in practical systems \cite{barz2010heralded}, we  assume that  Bob knows whether he has the assistance or not. % 

\begin{definition} 
\label{Definition:Capacity_Interception}
A $(2^{nR},2^{nR'},n)$   code with unreliable entanglement assistance consists of the following:
\begin{itemize}
\item
Two message sets $[1:2^{nR}]$ and $[1:2^{nR'}]$, % 
\item
a pure entangled state $\Psi_{G^{n}_A, G^{n}_B}$, 
\item
a collection of encoding maps $\{\mathcal{F}_{G^{n}_A\xrightarrow{}A^n}^{(m,m')}\}% 
$, % 
and 

\item
two % 
POVMs, $\mathcal{D}_{B^nG^{n}_B}= \{D_{m,m'}\}$ and $\mathcal{D}^{*}_{B^n}= \{D^{*}_{m}\}$.
\end{itemize}
\end{definition}

The % 
scheme is depicted in Figure~\ref{figure_switch_Interception}. Alice  holds % 
$G^{n}_A$.
She chooses two 
messages $m$ and $m'$,  % 
 encodes by % 
\begin{align}
\rho_{A^nG^{n}_{B}}^{m,m'} = (\mathcal{F}_{G^{n}_A\xrightarrow{}A^n}^{(m,m')}\otimes\mathrm{id})
(\Psi_{G^{n}_AG^{n}_B})
\label{encoding_messages}
\end{align}
and transmits $A^n$. % 
The channel output % 
is 
$% 
\rho_{B^n E^n G^{n}_B}^{m,m'} =(\mathcal{N}_{A\to B E}^{\otimes n}\otimes \mathrm{id})(\rho_{A^nG^{n}_B}^{m,m'}) % 
$. % 
Bob receives $B^n$.
Depending on the availability of the entanglement assistance, Bob decides whether to decode both messages or only one. If % 
$G^{n}_B$ is available, Bob performs % 
$\mathcal{D}_{B^nG^{n}_B}% 
$ to recover both messages. Otherwise, % 
Bob measures $\mathcal{D}^*_{B^n}% 
$ and estimates $m$ alone.

\begin{figure}[tbp]
\centering
        \includegraphics[scale=0.5, trim={3.5cm 9cm 0 10.25cm}, clip]{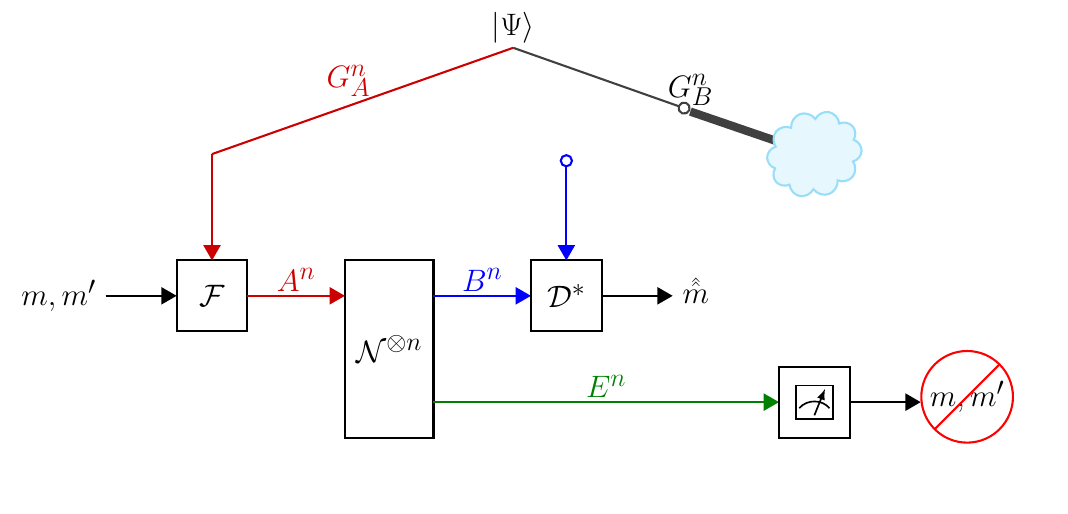} % 
\\ \vspace{-0.5cm}    
    \caption{Passive eavesdropper. % 
    The resource may get lost to the environment.
    }
    \label{figure_switch_Passive}
\end{figure}
We % 
have two maximum error criteria; in the presence of entanglement assistance: % 
\begin{align}
&P% 
_{e,\max}(\Psi,\mathcal{F},\mathcal{D}) 
=\max_{m,m'}\left[ 1 - 
\trace(D_{m,m'} \,
\rho_{B^n G_B^n}^{m,m'}
)
\right]
\,,
\nonumber
\intertext{and without % 
entanglement assistance:}% 
&P^{*% 
}
_{e,\max}(\Psi,\mathcal{F},\mathcal{D}^*) 
=\max_{m,m'}\left[ 1 - 
\trace(D_{m}^* \,
\rho_{B^n }^{m,m'}
)
\right]
\,.
\label{max_error}
\end{align}

\section{Semantic Security}
We consider two security settings.
\subsection{Security Under % 
Interception}
Suppose that Eve may steal the entanglement resource $G_B^n $. 
In the pessimistic case, Eve intercepts the entanglement resource, and Bob decodes without it.  In other words, Alice and \emph{Eve} share the entanglement, instead of Bob. See Figure~\ref{figure_switch_Interception}(b). 

Semantic security requires that Eve cannot gain any information on Alice's message, regardless of the message distribution.
Hence,  
the state of Eve's resources needs to be close to a \emph{constant state} that does not depend on  Alice's messages.
Formally, define the security level under interception, % 
with respect to 
a constant state $\theta_{E^n G_B^n}$, by % 
\begin{align}
\Delta_{\text{SI}}% 
(\Psi,\mathcal{F},\theta_{E^{n}G_{B}^{n}})= 
\max_{m,m'}
\frac{1}{2} \norm{\rho_{E^{n}G_{B}^{n}}^{m,m'} - \theta_{E^{n}G_{B}^{n}}
}_1
\,.
\label{indistin_definition_Interception}
\end{align}
Notice that we include the entangled resource $G^{n}_B$ in the indistinguishability criterion due to the pessimistic case above. % 

\begin{definition}% 
A $(2^{nR}, 2^{nR'}, n, \epsilon, \delta)$ code with unreliable entanglement assistance and semantic security under interception satisfies 
$% 
\max\left(P% 
_{e, \max}(\Psi,\mathcal{F} \,,\; \mathcal{D}),P^{*% 
}_{e, \max}(\Psi,\mathcal{F},\mathcal{D}^*)\right)\leq\epsilon % 
$, % 
and there exists $\theta_{E^n G_B^n}$ such that % 
$
\Delta_{\text{SI}}(\Psi,\mathcal{F},\theta_{E^n G_B^n})\leq\delta 
$. % 
A  rate pair $(R,R')$ is called achievable if $\forall$ % 
$\epsilon,\delta> 0$ and % 
large $n$, there is % 
a $(2^{nR}, 2^{nR'}, n, \epsilon,\delta)$ code.
The  capacity region $\mathcal{C}_{\text{SI}}(\mathcal{N})$ with unreliable entanglement assistance and semantic security under interception  is the closure of the set of all such pairs. 
\end{definition}

\subsection{% 
Passive Eavesdropper}
The passive model assumes that  Eve does not gain access to the % 
resource $G_B^n $. 
See Figure~\ref{figure_switch_Passive}. 
The security level % 
is now % 
\begin{align}
\Delta_{\text{PE}}(\Psi,\mathcal{F},\theta_{E^{n}})= 
\max_{m,m'}
\frac{1}{2} \norm{\rho_{E^{n}}^{m,m'} - \theta_{E^{n}}
}_1
\label{indistin_definition_Passive}
\end{align}
(cf. \eqref{indistin_definition_Interception}). % 
The % 
capacity region % 
$\mathcal{C}_{\text{PE}}(\mathcal{N})$ % 
is defined accordingly. % 

 \begin{remark}
 As opposed to % 
 \cite{Lederman2024Secure}, 
we do not assume that the messages are uniformly distributed. Instead, % 
\eqref{max_error}-\eqref{indistin_definition_Passive} % 
involve a maximum,
in accordance with 
the semantic security approach. % 
 \end{remark}

\begin{figure}[tbp]
\centering
        \includegraphics[scale=0.55, trim={4.5cm 8.65cm 0 8.5cm}, clip]{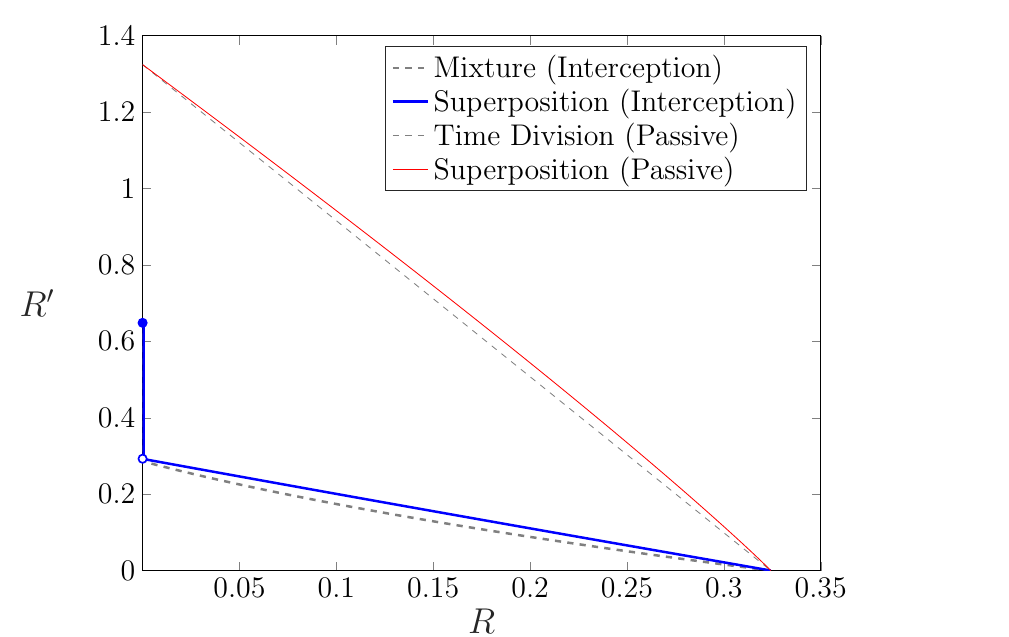} % 
        \\ % 
    \caption{Achievable rate regions for the amplitude damping channel with unreliable entanglement assistance and semantic security,  % 
    for $\gamma = 0.3$}
    \label{figure_Depolarizing_Example}
\end{figure}

\section{Main Results}% 
\label{Section:Results_Interception}

\subsection{Security Under Interception}
We consider  communication with unreliable entanglement assistance and semantic security. Recall that  Alice does not know whether the entanglement resource has reached Bob's location, hence  she encodes two messages, at rates $R$ and $R'$ (see 
Section~\ref{section_definitions_interception}). % 
If entanglement assistance is available to Bob, he recovers both messages. Yet, if Eve has stolen 
the resource, then he recovers the first message alone.
Let $\mathcal{N}_{A\to BE}$ be a quantum wiretap channel. Define
\begin{align}
&\mathcal{R}_\text{SI}(\mathcal{N})
\equiv
\bigcup_{ p_X, \varphi_{G_{1}G_{2}}, \mathcal{F}^{(x)} } 
\nonumber\\
&\left\{ \begin{array}{rl}
  (R,R') \,:\;
    R   \leq       [I(X;B)_\omega - I(X;EG_{2})_\omega]_+  \\
    R'  \leq        [I(G_{2};B|X)_\omega - I(G_{2};E|X)_\omega]_+ \\
	\end{array}
\right\} \ 
\label{R_SI}
\end{align}
where
\begin{align}
    \omega_{XG_{2}A}&\equiv\sum_{x\in\mathcal{X}}p_X(x)\ketbra{x}\otimes(\text{id}\otimes\mathcal{F}^{(x)}_{G_{1}\to A})(\varphi_{G_{2}G_{1}}) \,,
\nonumber\\
    \omega_{XG_{2}BE}&\equiv(\text{id}\otimes\text{id}\otimes\mathcal{N}_{A\to BE})(\omega_{XG_{2}A}) \,,
    \label{Equation:omegaXG2BE}
\end{align}
with $[x]_+\equiv \max(x,0)$.
Our main result on the interception model is given in the theorem below. % 
\begin{theorem}
\label{Theorem:Inner_Bound_Interception_Model}
The region $\mathcal{R}_\text{SI}(\mathcal{N})$ is achievable
 with unreliable entanglement assistance and semantic security under interception.
 That is,
the  capacity region % 
is bounded by
\begin{align}
\mathcal{C}_{\text{SI}}(\mathcal{N}) \supseteq
\mathcal{R}_\text{SI}(\mathcal{N}) \,.
\end{align}
\end{theorem}

The proof of Theorem \ref{Theorem:Inner_Bound_Interception_Model} is given in section \ref{Proof_Achievability}. 
Our proof modifies the methods of Cai \cite{cai2014maximum, pereg2023multiple}, originally applied to multiple-access channels (without secrecy), using random message permutations.

\subsection{% 
Passive Eavesdropper}
\label{Section:Results_Passive}
Here, the entanglement assistance  can be lost to the environment, beyond Eve's reach.
Define
\begin{align}
&\mathcal{R}_\text{PE}(\mathcal{N})
\equiv
\bigcup_{ p_X, \varphi_{G_{1}G_{2}}, \mathcal{F}^{(x)} } 
\nonumber\\ 
&\left\{ \begin{array}{rl}
  (R,R') \,:\;
    R   \leq    &    [I(X;B)_\omega - I(X;E)_\omega]_+  \\
    R'  \leq    &    I(G_{2};B|X)_\omega \\
	\end{array}
\right\} \ 
\label{R_PE}
\end{align}
where $\omega_{XG_{2}BE}$ is as in  \eqref{Equation:omegaXG2BE}.
Our main result on the passive model is given  below.
\begin{theorem}
\label{Theorem:Inner_Bound_Passive_Model}
The region $\mathcal{R}_\text{PE}(\mathcal{N})$ is achievable
 with unreliable entanglement assistance and a passive eavesdropper.
 That is,
the   capacity region  is bounded by
\begin{align}
\mathcal{C}_{\text{PE}}(\mathcal{N}) \supseteq
\mathcal{R}_\text{PE}(\mathcal{N}) \,.
\end{align}
\end{theorem}
 As the assistance remains secure from the eavesdropper, Alice and Bob can use the entanglement resources in order to generate a secret key.
 Alice can then apply the one-time pad encoding to the excess message $m'$. % 
 The rest of the analysis  % 
 follows similar steps as for % 
Theorem~\ref{Theorem:Inner_Bound_Interception_Model}.
The details are omitted.

\subsection{Example - Amplitude Damping Channel}
Consider the amplitude damping channel
$% 
    \mathcal{N}(\rho) =  K_0\rho K_0^{\dagger}+K_1\rho K_1^{\dagger}
$, % 
with
$% 
    K_0 = \ketbra{0}+\sqrt{1-\gamma}\ketbra{1} 
$ and
$
    K_1 = \sqrt{\gamma}\ketbra{0}{1}
$, % 
with $\gamma \in [0,1]$.
We numerically compute achievable regions for each setting, using the following ensemble. % 
Define % 
$% 
    \ket{u_\beta} = \sqrt{1-\beta}\ket{0}\otimes\ket{0}+\sqrt{\beta}\ket{\Phi}
$, % 
and set % 
$% 
    \ket{\phi_{G_1G_2}} = \frac{1}{\norm{u_\beta}}\ket{u_\beta}
$, % 
$
    p_{X} = \left(\frac{1}{2}, \frac{1}{2} \right)
$, and $
    \mathcal{F}^{(x)}(\rho) = \Sigma_X^x \rho \Sigma_X^x % 
$, $x\in\{0,1\}$, % 
where $\Sigma_X$ is the Pauli bit-flip operator.

The resulting achievable regions, for the interception and passive models, are indicated by the solid lines in % 
Figure~\ref{figure_Depolarizing_Example}, in blue and red, respectively. % 
For comparison, the dashed lines indicate the regions that are achieved through a classical mixture of optimal strategies, for communication with and without entanglement assistance.
In the interception model, time division is impossible because the use of entanglement can lead to a leakage of guaranteed information.
As can be seen in 
 Figure~\ref{figure_Depolarizing_Example}, the  point 
 $(R,R')=(0, 0.648)$ % 
 is disconnected from the set of
 boundary points for which $R>0$. 
In the passive model, on the other hand, we see that our coding scheme outperforms time division.

\section{Proof of Theorem~\ref{Theorem:Inner_Bound_Interception_Model}}
\label{Proof_Achievability}

We show achievability with semantic security under interception. The main technical novelty is  in Section~\ref{Subsection:Semantic_Analysis}.
We begin with useful results from our previous work  \cite{Lederman2024Secure}.

\subsection[Random Code Analysis (Previous Work)]{Random Code Analysis (Previous Work {\cite{Lederman2024Secure}})}
The code construction and previous results from \cite{Lederman2024Secure} are summarized below.
Fix % 
a pure state 
$\ket{\phi_{G_{1}G_{2}}}% 
$ and a collection of isometries % 
$\{ F^{(x)}_{G_1 \to A} \}$. % 
Denote
$% 
    \ket{\psi^{x}_{AG_2}} = (F_{G_{1}\to A}^{(x)}\otimes\identity)\ket{\phi_{G_{1}G_{2}}}
$ and % 
$
    \omega^{x}_{BEG_2} = (\mathcal{N}_{A\to BE}\otimes\text{id})(\psi^{x}_{AG_2})
$. % 
Suppose Alice and Bob would like to share $\ket{\phi_{G_{1}G_{2}}}^{\otimes n}$.

\vspace{0.1cm}
\emph{Classical Codebook Generation:} 
Select a random codebook % 
$% 
\mathscr{C}_1=\big\{x^{n}(m,k) \big\}_{
m\in [1:2^{nR}],% 
k\in [1:2^{nR_0}]}
$,
each codeword is i.i.d.  $\sim p_X$.
Denote the Heisenberg-Weyl operators of dimension $D$ by $\{\Sigma_{X}^{a}\Sigma_{Z}^{b}\}$. Consider a Schmidt decomposition $\ket{\psi^x_{AG_2}} = \sum_{y\in\mathcal{Y}}\sqrt{p_{Y|X}(y|x)}\ket{\xi_{y|x}}\otimes\ket{\xi^{'}_{y|x}}$. For every conditional type class $\mathcal{T}_n(t|x^n)$ in $\mathcal{Y}^n$, define
$% 
    V_{t}(a_t,b_t,c_t) = (-1)^{c_t}\Sigma_{X}^{a_t}\Sigma_{Z}^{b_t}
$, % 
for $a_t, b_t \in \{0,...,\abs{\mathcal{T}_n(t|x^n)}-1\}$, $c_t\in \{0,1\}$. Consider $U(\gamma) = \bigoplus_{t}V_{t}(a_t,b_t,c_t)$ with $\gamma = ((a_t, b_t, c_t)_t)$,
and let $\Gamma_{x^{n}}$ denote the set of all such vectors $\gamma$.
Then, for every $m$ and $k$, select $2^{n(R'+R_0')}$ conditionally independent sequences,
$% 
\mathscr{C}_2=\big\{ \gamma(m',k'|x^{n}(m,k)\big\}_{% 
m'\in [1:2^{nR'}],% 
k'\in [1:2^{nR'_0}]}
$, % 
uniformly at random. % 
The codebook
$\mathscr{C}=(\mathscr{C}_1,\mathscr{C}_2)$
is publicly revealed. % 

\vspace{0.1cm}
\emph{Encoder:} 
Select $k$ and $k'$ uniformly at random.
To encode % 
$(m,m')$, apply $\bigotimes_{i=1}^n F_{G_1\to A}^{(x_i)}$ 
and then  $U(\gamma)$, with $x^n\equiv x^n(m,k)$ and
$\gamma\equiv \gamma(m',k'|x^n(m,k))$. % 
Transmit $A^n$. 

Denote the % 
output state by 
$% 
    \rho^{\gamma,x^{n}}_{B^{n}E^{n}G_2^{n}} % 
$. % 

\vspace{0.1cm}
\emph{Decoder:}   
Based on  \cite{pereg2023communication}, Bob can decode the guaranteed and excess messages such that 
the expected message-average error probabilities 
satisfy
\begin{align}
\mathbb{E} \left[\frac{1}{2^{n(R+R')}}\sum_{m,m'}
P_e(\mathscr{C}|m,m') \right]
&\leq \epsilon_1
\,,
\label{average_prob_1}
\\
\mathbb{E} \left[\frac{1}{2^{n(R+R')}}\sum_{m,m'}
P_e^*(\mathscr{C}|m,m') \right]
&\leq \epsilon_1
\,,
\label{average_prob_2}
\end{align}
provided that 
$% 
    R+R_0 < I(X;B)_\omega - \epsilon_2 % 
$ and 
$
       R'+R'_0 < I(G_{2};B|X)_\omega -\epsilon_3 % 
$. % 

We move to the security level. 
Denote 
\begin{align}
    \Delta_{m'|m,k}(\mathscr{C}) &= \frac{1}{2}\norm{\frac{1}{2^{nR'_0}}\sum_{k'=1}^{2^{nR'_0}}\rho_{E^{n}G_{2}^{n}}^{\gamma(m',k'|x^n), x^{n}} - \zeta^{x^n}_{E^{n}G_{2}^{n}}
    }_1
\,,
\nonumber
\\
    \Delta_{m}^*(\mathscr{C}) &= \frac{1}{2}\norm{\frac{1}{2^{nR_0}}\sum_{k=1}^{2^{nR_0}}\omega^{x^n}_{E^{n}G_{2}^{n}}- \omega_{E G_{2}}^{\otimes n}}_1
\,,    
\label{Equation:Delta*}
\end{align}
with $x^n\equiv x^n(m,k)$, and $\zeta^{x^n}_{E^{n}G_{2}^{n}} = \frac{1}{\abs{\Gamma_{x^n}}} \sum_{\gamma\in\Gamma_{x^n}}\rho^{\gamma, x^n}_{E^{n}G_{2}^{n}}$.

Based on the quantum covering lemma \cite{ahlswede2002strong}, we have the following indistinguishability bounds: \cite{Lederman2024Secure}
 \begin{align}
     \Pr\left(
     \Delta_{m}^*(\mathscr{C})
     >e^{-\frac{\lambda }{2}n} \right)
     &\leq % 
     \exp{-% 
     2^{n(R_0-I(X;EG_2)_\omega-\epsilon_4)}} \, % 
\nonumber
\intertext{$\forall m$, and}
     \Pr\left(
     \Delta_{m'|m,k}(\mathscr{C})
     > e^{-\frac{\mu }{2} n} \right)
     &
     \leq 
     \exp{-% 
     2^{n(R'_0-I(G_2;E|X)_\omega% 
     -\epsilon_{5})}} 
     \label{prob_sec_3}
 \end{align}
 $\forall (m,k,m')$, 
given $x^n\equiv x^n(m,k)$. % 
The last two bounds tend to zero in a double exponential rate for
 $% 
     R_0 = I(X;EG_{2})_\omega +2\epsilon_{4} \,.
 $ and % 
$% 
     R'_0 = I(E;G_{2}|X)_\omega +2\epsilon_{5} % 
$. % 

\subsection{De-randomization}
We now show that there exists a deterministic codebook under the requirements of average error probabilities and maximal indistinguishability. 
Consider the following  error events,
\begin{align}
    \mathcal{A}_1 &= \{\frac{1}{2^{n(R+R')}}\sum_{m,m'}
P_e(\mathscr{C}|m,m')>\sqrt{\epsilon_1}\} \,,
\\
    \mathcal{A}_2 &= \{\frac{1}{2^{n(R+R')}}\sum_{m,m'}
P_e^*(\mathscr{C}|m,m')>\sqrt{\epsilon_1}\} \,,
\\
     \mathcal{B} &=\{ 
     \exists (m,m'): \frac{1}{2} \norm{\rho_{E^{n}G_{B}^{n}}^{m,m'} - \omega_{E G_{2}}^{\otimes n}}_1
     > \delta \} \,.
    \label{indistinguishability_demand}
\end{align}
By the union bound,
\begin{align}
    \Pr(\mathcal{A}_0\cup \mathcal{A}_1 \cup \mathcal{B}) \leq
    \Pr(\mathcal{A}_0)+\Pr(\mathcal{A}_1)+\Pr(\mathcal{B}) \,.
    \label{prob_union_bound}
\end{align}
By Markov's inequality, % 
$\Pr(\mathcal{A}_j)\leq \sqrt{\epsilon_1}$ (see  \eqref{average_prob_1}-\eqref{average_prob_2}).
As for the last term, by the triangle inequality, % 
\begin{align}
    &\frac{1}{2} \norm{\rho_{E^{n}G_{B}^{n}}^{m,m'} - \omega_{E G_{2}}^{\otimes n}}_1 
    \nonumber\\
    &= \frac{1}{2} \norm{\frac{1}{2^{n(R_0+R'_0)}}\sum_{k=1}^{2^{nR_0}}\sum_{k'=1}^{2^{nR'_0}}\rho_{E^{n}G_{B}^{n}}^{\gamma(m',k'|x^n), x^{n}} - \omega_{E G_{2}}^{\otimes n}}_1
    \nonumber \\
    &\leq \frac{1}{2} \norm{\frac{1}{2^{nR_0}}\sum_{k=1}^{2^{nR_0}} \left(\frac{1}{2^{nR'_0}}\sum_{k'=1}^{2^{nR'_0}}\rho_{E^{n}G_{B}^{n}}^{\gamma(m',k'|x^n), x^{n}} 
    - \zeta^{x^n}_{E^{n}G_{2}^{n}}
    \right)}_1
    \nonumber\\
    &
    + \frac{1}{2} \norm{\frac{1}{2^{nR_0}}\sum_{k=1}^{2^{nR_0}}\zeta^{x^n}_{E^{n}G_{2}^{n}}
    - \omega_{E G_{2}}^{\otimes n}}_1
    \nonumber\\
    &\leq \frac{1}{2^{nR_0}}\sum_{k=1}^{2^{nR_0}} \Delta_{m'|m,k}(\mathscr{C})
    \nonumber\\
    &+  \frac{1}{2} \norm{\frac{1}{2^{nR_0}}\sum_{k=1}^{2^{nR_0}}\zeta^{x^n(m,k)}_{E^{n}G_{2}^{n}}
    - \omega_{E G_{2}}^{\otimes n}}_1 \,.
\end{align}
If we were to remove the encoding of $\gamma$, % 
then Eve's output would have been
$\omega^{x^n}_{E^{n}G_2^{n}} $, instead of $\zeta^{x}_{E^{n}G_{2}^{n}}$. 
Therefore, by trace monotonicity  under quantum operations \cite[Ex. 9.1.9]{Wilde:17b},
the last trace norm is bounded by 
$\Delta_{m}^*(\mathscr{C}) $ (see \eqref{Equation:Delta*}).
Thus, 
\begin{align}
    &%\Pr(\mathcal{B})
    %=
    \Pr(\frac{1}{2} \norm{\rho_{E^{n}G_{B}^{n}}^{m,m'} - \omega_{E G_{2}}^{\otimes n}}_1 > \delta) 
    \nonumber\\
    &
    \leq \Pr(\frac{1}{2^{nR_0}}\sum_{k=1}^{2^{nR_0}}\Delta_{m'|m,k}(\mathscr{C}) > \frac{\delta}{2}) + \Pr(\Delta_{m}^*(\mathscr{C}) > \frac{\delta}{2})
    \nonumber\\
    &\leq \Pr(\exists k : \Delta_{m'|m,k}(\mathscr{C}) > \frac{\delta}{2}) + 
    \exp\left\{-2^{n\epsilon_4} \right\}
    \nonumber\\
    &\leq \exp\left\{-2^{n\epsilon_6}\right\}
\end{align}
for some $\epsilon_6>0$ and sufficiently large $n$.
Then, it follows that 
\begin{align}
\Pr(\mathcal{B})&\leq 
2^{n(R+R_0)}\cdot \exp\left\{-2^{n\epsilon_6}\right\}
\\
&\leq 
\exp\left\{-2^{\frac{1}{2}n\epsilon_6}\right\}
\end{align}
for large $n$.
We deduce that there exists a deterministic codebook
$\mathscr{C}$ % 
such  that the message-average error % 
and indistinguishability tend to zero.

\subsection{Semantic Security}
\label{Subsection:Semantic_Analysis}
We now complete the analysis for the maximum criteria.  
The proof modifies the methods of Cai \cite{cai2014maximum, pereg2023multiple}, originally applied to multiple-access channels.

\subsubsection{Guaranteed information (expurgation)}
Consider  the semi-average error probability, % 
\begin{align}
    e({m}) \equiv \frac{1}{2^{nR'}}\sum_{m'=1}^{2^{nR'}}
    P_e(\mathscr{C}|m,m') \,.
\end{align}
Based on the analysis above, the average of 
$\{e(m)\}_{m=1}^{2^{nR}}$ is bounded by 
$\epsilon_1^{1/2}$.
Therefore,
at most a fraction of $\lambda = {\epsilon_1}^{1/4}$ of the messages $m$ have  
$e({m})>\lambda$.
Then, we can expurgate the worst $\lambda\cdot 2^{nR}$ messages, and the corresponding codewords.
The guaranteed rate of the expurgated code is
$ R-\frac{1}{n}\log((1-\lambda)^{-1})$, which tends to $R$ as 
$n
\to\infty$.
Denote the expurgated message set by $\mathcal{M}$.

\subsubsection{Excess information (message permutation)}
We now construct a new code to satisfy the maximum criteria. 
The transmission consists of two stages.
In the first stage, Alice selects a uniform ``key" $L\in [1:n^2]$.
Assuming that $R'>0$, Alice can send $L$ with negligible rate loss, such that the message-average error probabilities vanish.
In the second stage, Alice chooses a permutation $\pi_{L}$ on the message set $[1:2^{nR'}]$, and encodes the message pair $(m_0,m_0')=(m,\pi_{L}(m'))$ using the codebook $\mathscr{C}$.
Bob obtains an estimate,
$\hat{L}$ % 
and 
$(\hat{m}_0,\hat{m}_0')$, and then declares his estimation for the original messages as 
$\hat{m}=\hat{m}_0$ and $\hat{m}'=\pi_{\hat{L}}^{-1}(\hat{m}_0')$.

Based on our previous analysis, the message-average error probability in the first stage is bounded by 
$% 
    \Pr(\hat{L} \neq L) = \frac{1}{n^2}\sum_{\ell=1}^{n^2}P_{e}(\mathscr{C}|1,\ell) \leq \sqrt{\epsilon_1}
$. % 
Now, consider the second block.
Let 
$\Pi_1,\ldots,\Pi_{n^2}$ be an i.i.d. sequence of random permutations, each uniformly distributed on the permutation group on the excess message set
$[1:2^{nR'}]$.
Denote the associated random codebook by 
$\Pi(\mathscr{C})$.
Then, for a given $m'$,
\begin{align}
    \Pr(\Pi_{\ell'}(m') = \bar{m}')= \frac{(2^{nR'}-1)!}{(2^{nR'})!} = \frac{1}{2^{nR'}}
\end{align}
for all $\bar{m}'\in [1:2^{nR'}]$ and $\ell'\in 
[1:n^2]$.
Thus, for every message pair
$(m,m')\in \mathcal{M}\times [1:2^{nR'}]$,
\begin{align}
    &\mathbb{E}\left[ P_{e}^{(n)}(\Pi(\mathscr{C})|m,m') \right] 
    \nonumber\\
    &= \sum_{\bar{m}'} \Pr(\Pi_{\ell'}(m')=\bar{m}') P_{e}^{(n)}(\mathscr{C}|m, \bar{m}') 
    \nonumber\\
    &= \frac{1}{2^{nR'}}\sum_{\bar{m}'}P_{e}^{(n)}(\mathscr{C}|m, \bar{m}') = {e}(m) \leq \lambda
    \,.
\end{align}

Now, by the Chernoff bound  \cite[Lemma 3.1]{cai2014maximum}, % 
\begin{align}
    \Pr\left(\frac{1}{n^2}\sum_{l'=1}^{n^2}P_{e}^{(n)}(\Pi(\mathscr{C})|m.m') > 4\lambda\right) < e^{-\lambda  n^2}
\end{align}
Therefore, the probability that, for some $(m,m')$,
$% 
    \frac{1}{n^2}\sum_{l'=1}^{n^2}P_{e}^{(n)}(\Pi(\mathscr{C})|m,m') > 4\lambda % 
$, % 
tends to zero in a super-exponential rate by the union bound.
We deduce that there exists a realization 
$(\pi_1,\ldots,\pi_{n^2})$ such that
\begin{align}
    P_{e}^{(n)}(\pi(\mathscr{C})|m,m') 
    =\frac{1}{n^2}\sum_{\ell'=1}^{n^2}P_{e}^{(n)}(\pi_{\ell'}(\mathscr{C})|m,m')
    \leq 4\lambda
    \label{Error_Probability_Semantic}
\end{align}
for all $(m,m')\in \mathcal{M}\times [1:2^{nR'}]$.
\qed

\section{Summary and Discussion}
\label{Section:Discussion}
We consider semantic security % 
with unreliable entanglement assistance, due to one of two reasons: Interception or loss. In the interception model, Eve may steal the entanglement assistance
(see Figure~\ref{figure_switch_Interception}(b)). Whereas, loss implies that Eve is passive and the assistance may get lost to the environment (see Figure~\ref{figure_switch_Passive}).  
 The authors have recently considered the interception model with
 strong secrecy \cite{Lederman2024Secure}, 
 assuming that the information is uniformly distributed.

Here, we derive
 achievable rates for both the interception and loss models,  subject to a maximal error criterion and semantic security. % 
 In the interception model, the guaranteed rate bound includes both Eve's system $E$  and Bob's entangled resource $G_B$ (see \eqref{R_SI}), which 
 reflects  Eve's access to the entanglement assistance if she succeeds to intercept the resource. % 
 On the other hand,  in the passive eavesdropper model, the guaranteed rate bound does not involve the entangled resource $G_B$ (see  \eqref{R_PE}), as the assistance is % 
 beyond Eve's reach.

Moreover, the bound on the excess rate, in the passive model,    does not include Eve's system % 
at all (see \eqref{R_PE}), i.e.,
secrecy does not entail a rate reduction. % 
This is expected because given reliable entanglement assistance, Alice and Bob can secure a shared key, and apply the one-time pad encryption to   the excess message. 

As an example, we consider the amplitude damping channel. % 
In the interception model, where Eve can actively intercept the entanglement assistance, time division impossible and the boundary of our achievable region is disconnected. 
This occurs % 
since interception can severely impact the achievable rates. % 
In the passive model, on the other hand, our encoding scheme outperforms time division.

Some questions still remain open, as 
we do not have a full understanding of the behavior of the capacity region, its convexity properties, and the type of entanglement that allows positive guaranteed rate under interception.
Furthermore, while our previous work \cite{Lederman2024Secure} has presented a regularized characterization for the special class of degraded channels,  a single-letter capacity formula in special cases could lead to further insights.

\section*{Acknowledgments}
U. Pereg and M. Lederman were supported by  ISF % 
Grants n. 939/23 and 2691/23, DIP % 
n.
2032991, and  Nevet Program of the Helen Diller Quantum Center at the Technion, n. 	2033613.
U. Pereg was also supported by the Israel VATAT % 
Program for Quantum Science and Technology % 
n. 86636903, and the Chaya Career Advancement Chair, n. % 
8776026.

\renewcommand{\theequation}{\thesection.\arabic{equation}}
\begin{appendices} % 
{

}
\end{appendices}

\balance
\bibliography{references}{}

\end{document}